\documentclass{vgtc}                         

\graphicspath{{figures/}{pictures/}{images/}{./}}

\newcommand\tb[1]{\textcolor{blue}{#1}}
\renewcommand\tb[1]{#1}

\usepackage{times}                     
      
\usepackage{mathptmx}

\onlineid{1083}
\vgtccategory{Technique or Algorithm}

\vgtcinsertpkg

\title{Chronoblox: Chronophotographic Sequential Graph Visualization}

\author{Quentin Lobbé\textsuperscript{1}
	\and Camille Roth\textsuperscript{1,2} 
	\and Lena Mangold\textsuperscript{1,2}
}
\affiliation{\scriptsize \textsuperscript{1}Computational Social Science Team, Marc Bloch Center, Berlin \\
\textsuperscript{2}Centre d'Analyse et de Mathématiques Sociales (CAMS), CNRS/EHESS, Paris }

\teaser{
	\centering
	\includegraphics[width=0.9\linewidth]{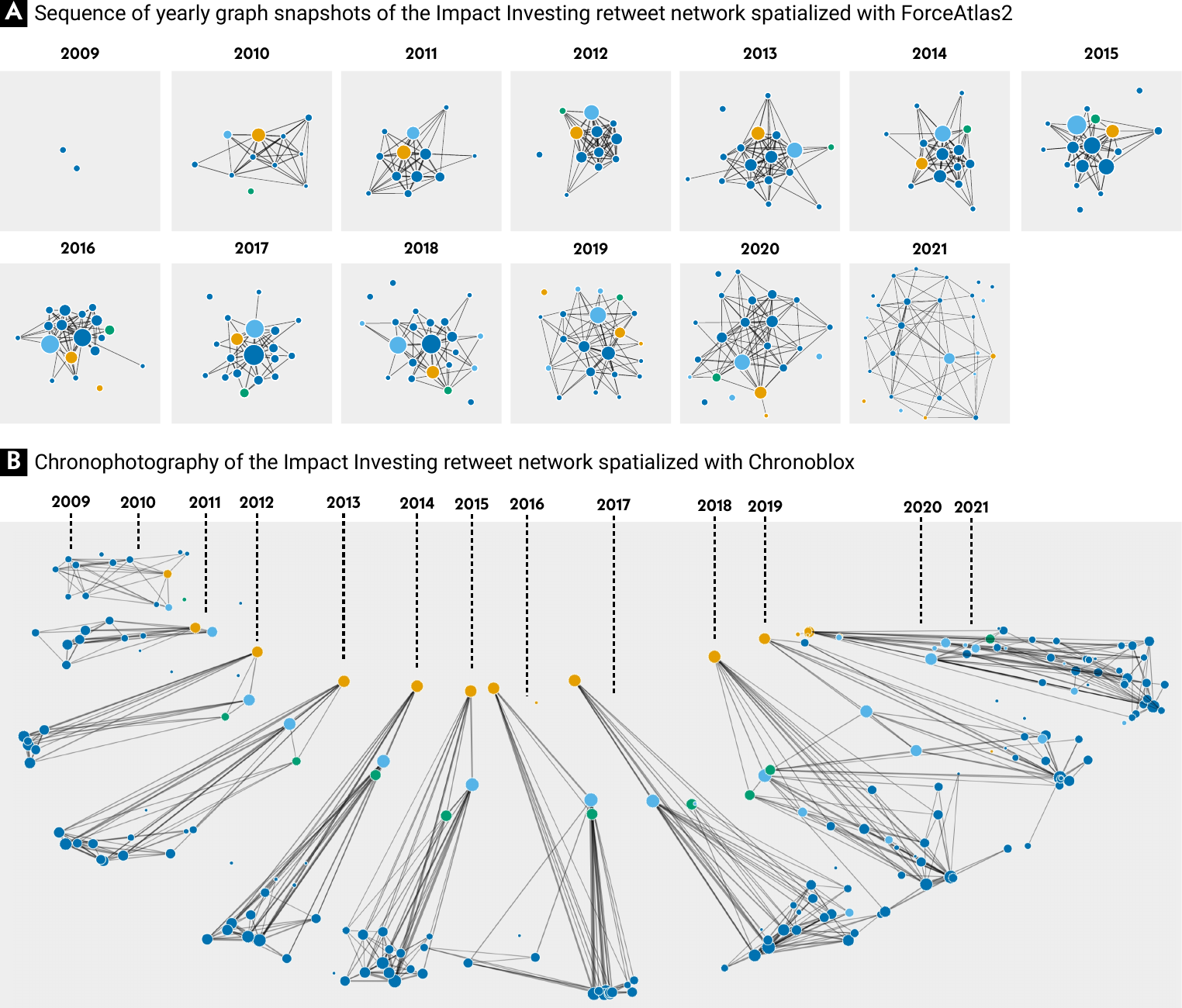}
	\caption{\textbf{Chronoblox} performs the chronophotography of a sequence of graphs (\textbf{B}). By using a single inter-temporal embedding space, Chronoblox lays out each snapshot in relation with the \tb{preceding or following ones}. The motion of the whole graph sequence can thus be interpreted in terms of micro to meso structural evolution. Both the temporal sequence (\textbf{A}) and its chronophotography (\textbf{B}) have been constructed from a retweet network where nodes are clusters of Twitter/X users detected via the Louvain algorithm (see~\autoref{application}). {Colors represent the most frequent home country of each node group}. Explore the \href{https://lobbeque.github.io/chronoblox_examples/impact_investing_louvain.html}{interactive version}.}
	\label{fig:teaser}
}

\abstract{
	We introduce Chronoblox, a system for visualizing dynamic graphs. Chronoblox consists of a chronophotography of a sequence of graph snapshots based on a single embedding space common to all time periods. The goal of Chronoblox is to project all snapshots onto a common visualization space so as to represent both local and global dynamics at a glance. In this short paper, we review both the embedding and spatialization strategies. We then explain the way in which Chronoblox translates micro to meso structural evolution visually. We finally evaluate our approach using a synthetic network before \tb{illustrating} it on a real world retweet network.
} 

\keywords{Dynamic graph visualisation, chronophotography, inter-temporal embedding, dimension reduction.}

\begin{document}
\maketitle

\section{Introduction}
\label{introduction}

Dynamic Graph Visualization can be categorized into two main streams, depending on whether the evolution of a sequence of graphs is depicted at the micro-level of nodes and links~\cite{beck2017}, or at the meso-level of node groups (such as clusters, communities, blocks)~\cite{vehlow2015state}. In either case, layouts integrate an explicit longitudinal dimension to avoid losing the sense of the inter-temporal connection between the elements of the various phases, as would, for instance, be the case with a simple series of graph snapshots \tb{trivially displayed one after the other}  (as in \autoref{fig:teaser}A).

These layouts, however, all have limitations when it comes to jointly grasping graph dynamics locally (from one phase to the other) and globally (over the whole sequence of graphs). Limitations also apply in terms of spatial scale. At the micro level, animation layouts succeed in revealing {local changes~\cite{ahn2011temporal} or simple transitions between phases~\cite{bach2013graphdiaries}}, but are by design hardly able to render global processes {without using a summary view in isolation~\cite{cakmak2020multiscale}}. At the micro-to-meso-level, superimposed layouts translate long-term evolution by stacking subsequent phases on top of each others~\cite{brandes2003visual,dwyer2002visualising}. However, they remain very sensitive to structural changes, making it difficult to scale up: they implicitly rely on the assumption that there is a sizable proportion of stable nodes across phases, which serve as fixed anchors over the graph sequence. Absent this stability, layout alignment across phases becomes an issue. {Although they fall outside the node-link scope of our paper, we remark that some matrix-based approaches have been able to solve this issue by following a piling metaphor~\cite{bach2014visualizing,bach2015small}.}
At the meso-level, so-called alluvial layouts represent a frequent strategy. They have made the choice to go without the \emph{intra-temporal} structure of each graph so as to focus instead on \emph{inter-temporal} streams of nodes~\cite{reda2011visualizing,rosvall2010mapping}. Some hybrid techniques yet try to enrich alluvial charts with topological information by displaying parts of the sequence of graphs alongside~\cite{mall2015netgram,vehlow2015visualizing}. But as these approaches lay out each individual phase into an independent visualization space, they induce discontinuities such as scaling inconsistencies, spurious shifts or meaningless rotations which may confuse the analysis.

Our proposal aims at uniting the various scales, both spatial (micro and meso) and sequential (local and global). More precisely, we argue 
that the movement of the entire sequence of graphs can be fully and concurrently integrated into a unified layout whose space meaningfully reflects the inter-temporal similarities of graphs across phases. Put differently, our goal is to project all the single phases onto a common visualization space so as to represent both local and global dynamics at a glance, thus making it possible to empirically interpret the whole sequence in terms of micro and meso-structural evolution. 

To do so, we have been inspired by a photographic technique called chronophotography: a process that aims at revealing the successive phases of an object in motion. A set of snapshots are first taken from a fixed spot before being printed on the same piece of film. The way in which each snapshot is laid out in relation to the previous/next ones enables the viewer to study the evolution of the observed object through time and space. With this in mind, we introduce Chronoblox, a novel layout designed to perform the chronophotography of a sequence of graphs, shown in \autoref{fig:teaser}B.

%
%

\section{Chronoblox}

Our network visualization proposal consists of a chronophotography of a sequence of graphs based on a single embedding space common to all time periods. This admittedly requires an embedding principle likely to reflect the main features of the temporal evolution of the graph. This, in turn, requires that nodes are labeled with metadata proper to each phase which, as we shall see, is always possible, irrespective of whether the input data is originally equipped with metadata of its own or not.

\subsection{Metadata Definition}
\label{chronoblox_sequence}

We first and foremost consider the standard aim of visualizing the evolution of the meso-level structure of a sequence of graphs $(G_1, ..., G_n)$ representing $n$ phases. The structure of each graph $G_t=(V_t,E_t)$ at phase $t$ may typically be depicted by gathering nodes into groups (for instance, via cluster detection~\cite{blondel2008fast} or stochastic blockmodeling~\cite{peixoto2019bayesian}). This makes it possible to build a sequence of meta-graphs $(\Gamma_1, ..., \Gamma_n)$ describing the level of node groups and their connections. More formally, applying a node grouping method at each phase $t$ yields a meta-graph $\Gamma_t=(B_t,\epsilon_t)$ where $B_t\subseteq\mathcal{P}(V_t)$ is the set of node groups and $\epsilon_t$ can be defined by aggregating edges of $E_t$ over node groups: for instance, by trivially merging edges between nodes of a group pair $(b,b')\in {B_t}^2$ i.e., $E_t\cap(b\times b')$. Now, each node $b\in B_t$ of the meta-graph may be labeled by the set of nodes of $V_t$ that it contains. In other words, the metadata associated with the node group represented by $b$ is straightforwardly the corresponding subset of $V_t$.

This focus on node groups, or possibly blocks, led us to name this approach ``Chronoblox''. Without loss of generality, note that we may also apply it to represent the evolution of the micro-level structure of the graph sequence $(G_1, ..., G_n)$ if natively equipped with metadata evolving through time --- for instance, if the data describes the evolving set of interests of scientists in a longitudinal co-authorship network, or of hashtags in a social media context.

To summarize, a crucial element of Chronoblox is that each node~$v$, of any of the graph sequence to be represented, may be labeled with metadata 
denoted as $m(v)$. 

\subsection{Common Embedding Space}
\label{chronoblox_embedding}

\begin{figure*}
	\centering
	\includegraphics[width=0.9\textwidth]{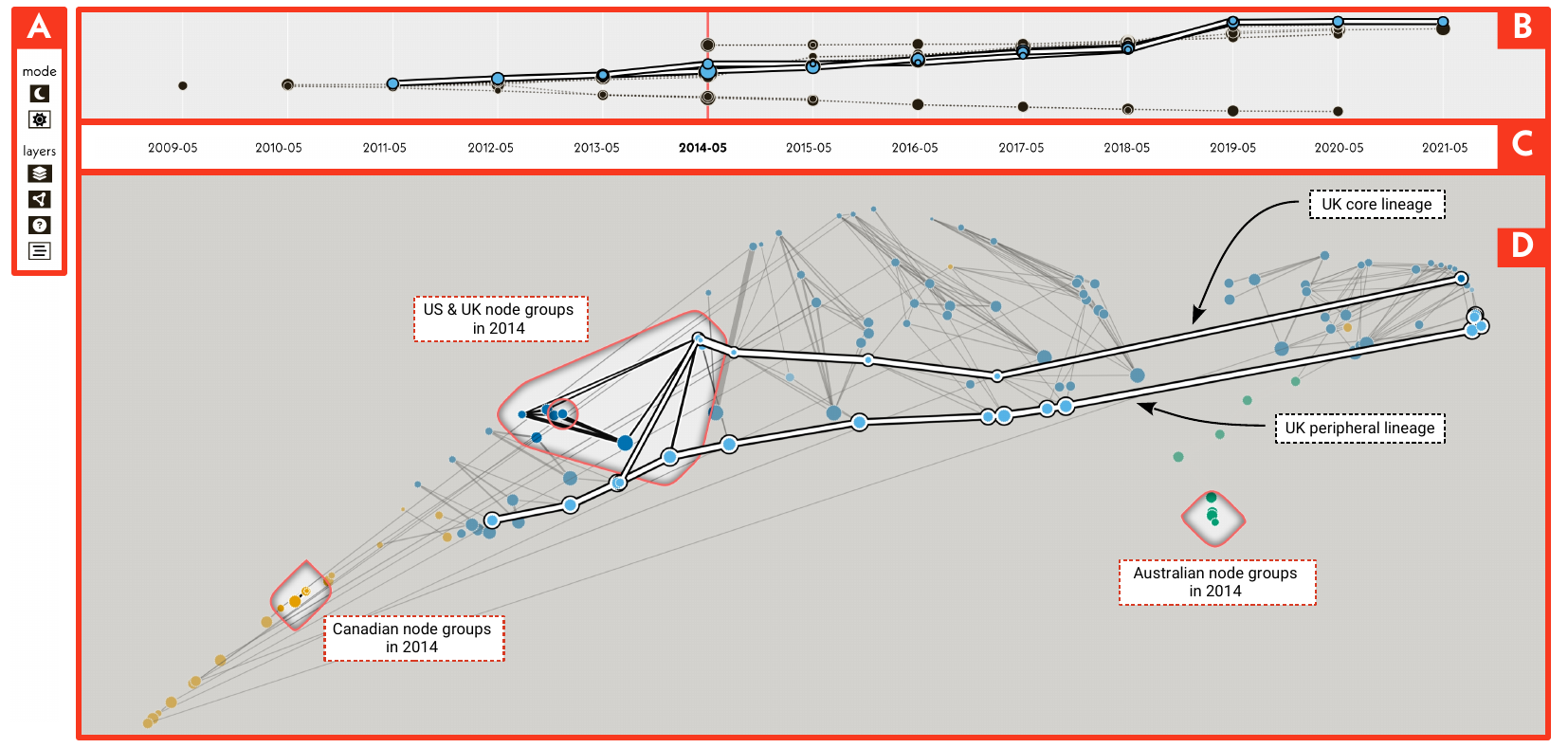}\vspace{-.9em}
	\caption{\textbf{The Chronoblox interface.} (A) Layer tab, (B) Alluvial view, (C) Timeline, and (D) Chronophotographic view. It shows the evolution of $\Xi$, the \textit{Impact Investing} retweet network with node groups computed, here as stochastic blocks of Twitter/X users (\href{https://lobbeque.github.io/chronoblox_examples/impact_investing_sbm.html}{see the interactive~version}). For illustrative purposes we circle country groups in 2014 and UK lineages over the whole sequence, see text in Sec.~\ref{application}.\vspace{-.2em}}
	\label{fig:interface}
\end{figure*}

The key idea of Chronoblox consists in fitting within a common embedding space all nodes across all phases, and thus all graphs. By doing so, we ensure that the spatialization of each phase occurs in relation to the whole sequence. {We are thus following in the footsteps of the emerging field of inter-temporal embedding layouts. Unlike existing works however, the subsequent phases will not be reduced to simple points~\cite{van2015reducing} or hidden behind summary views~\cite{cakmak2020multiscale}. With Chronoblox, the intra-temporal structure of each phase will always be visible in order to figure the micro to meso-level movements of the whole sequence of graphs.}

In the context of this short paper we practically restrain ourselves to the evolution of meta-graphs of node groups. In this respect, we adopt a processual view of social structures~\cite{abbott1995things} and consider the way in which node groups change at the meso-scale to bear strong explanatory power with regard to the overall network evolution.

As said above, each node $b$ of each meta-graph $\Gamma_t$ represents a node group of the corresponding graph $G_t$ and is labelled as such: $m_t(b)\subseteq V_t$. We then compute the similarity of each pair $b,b'\in\cup_{t\in\{1,..,n\}}B_t$ by means of a classical Jaccard index $|m(b)\cap m(b')|/|m(b)\cup m(b')|$, subsequently populating a matrix $M$ of similarities between all node groups encountered at any phase. Embedding $S$ into a common space implies that similar node groups shall be located in similar areas, thus ensuring an inter-temporal consistency of the graph spatializations of each phase --- graphs exhibiting similar node groupings shall be positioned closeby to one another and, likewise, significant structural transitions shall result in the corresponding graphs being located in different areas. In practice, we construct a 64-dimensional embedding space from $S$ via the Node2Vec algorithm~\cite{grover2016node2vec}. Random walks generate sentences from $M$ (i.e. lists of node groups) that are later used to calculate an embedding vector for each node group of each phase.

\subsection{Chronophotographic Projection}
\label{chronoblox_projection}

The 64-dimensional space wherein nodes are inter-temporally embedded eventually requires a two-dimensional projection before visualization. While dimension reduction methods are good at revealing salient cluster structures and distributional features, respecting both the local and global structure of high-dimensional objects has long been challenging~\cite{wattenberg2016use}. Linear approaches (such as PCA or MDS) struggle to jointly preserve both raw local distances and graph structure, whereas non-linear approaches (such as t-SNE or UMAP) are prone to create spurious clusters of points at a global scale. The recent PaCMAP algorithm~\cite{wang2021understanding} appears to solve this dilemma by focusing on mid-near pairs of data in order to preserve local and global properties. We deem this feature important to organize the chronophotographic spatialization at both scales --- in particular, to preserve the notion that, in this reduced visualization space translation or rotation can figure meso-scale changes while overlap can be interpreted as structural stability.

We here use the default PaCMAP parameters and obtain pairs of coordinates attached to each node group $b$ at any phase. Note that as PaCMAP introduces some randomness during its initialization step, distinct chronophotographic projections are likely to be produced upon each run; nonetheless each projection should by construction similarly satisfy the above-mentioned constraints and properties. 

%
%

\section{The Chronoblox Interface}
\label{interface}

We now present the Chronoblox interface where users can explore and interact with chronophotographic projections of sequence of meta-graphs $(\Gamma_1, ..., \Gamma_n)$, where we highlight three main features:\footnote{The source code of both the Chronoblox layout and interface are available at \url{https://anonymous.4open.science/r/chronoblox-BE82}.}\\

\noindent-~\textbf{Chronophotographic View.} This is the main view (\autoref{fig:interface}D) where node groups $B_t$ are represented by circles and placed using the PaCMAP coordinates. Intra-temporal edges $E_t$ connect these groups with solid black lines.  The focus on a given phase and meta-graph $\Gamma_t$ is visible on the timeline (\autoref{fig:interface}C) and moves to an adjacent phase $t-1$ or $t+1$ using arrow keys. The meta-graph in focus is painted in black and highlighted with red convex hulls {(we use a state of the art standard~\cite{bach2013graphdiaries})}.  Each meta-graph $\Gamma_t$ is slightly transparent to underscore overlaps with adjacent phases.\\

\noindent-~\textbf{Inter-Temporal Lineages.} The matrix $M$ of similarities provides inter-temporal similarities of node groups across periods. In particular, it makes it possible to relate node groups with similar node groups of immediately preceding and following phases, which denote parent/children lineage relationships across phases. Going further, and inspired by previous works which {have introduced a notion of inter-temporal lineage~\cite{palla2007quantifying,chavalarias2013phylomemetic}, we enrich the interface by modeling such lineages} as connected components in the network of lineage relationships. In~\autoref{fig:interface}B and ~\autoref{fig:interface}D, node groups belonging to the same connected component are circled with a dashed stroke, and connected through a solid white line; clicking on a node group reveals the lineage it belongs to.\footnote{For readability, inter-temporal similarity links have first been filtered by means of a symmetrical Herfindahl-Hirschman Index~\cite{rhoades1993herfindahl}, which helps focus on the most important links while removing the comparatively negligible links, both from the perspective of one node and from the perspective of the other (hence the symmetrical).}

\noindent-~\textbf{Alluvial View.} Chronoblox endeavors at spatializing similar node groups nearby --- accordingly, sequences of similarly groups are likely to follow visually consistent curvilinear trajectories. We contend that this property provides the basis for an optimized alluvial chart for free. In effect, reducing further the PaCMAP coordinates to unidimensional coordinates via PCA which should not vary too much from a phase to the other. The alluvial view of \autoref{fig:interface}B illustrates just that, by vertically placing node groups according to this single coordinate, while the horizontal axis represents time. It also features the inter-temporal similarity links connecting pairs of node groups. Node groups and lineages appear to be naturally sorted so as to minimise inter-temporal link crossings.



%
%

\section{Evaluation}
\label{evaluation}

The field of graph node grouping methods not only helps to build the metadata necessary to Chronoblox in the meta-graph case, but also to design a controlled environment against which to benchmark it. This field includes two main types of approaches~\cite{peixoto2021descriptive}. First, descriptive methods often aim at describing node groupings according to context-dependent partition-optimization criteria, such as modularity in the so-called Louvain algorithm~\cite{blondel2008fast} which principally uncovers cohesive node groupings or \emph{clusters}. Second, inferential techniques posit a generative process responsible for structure of the observed graph, by fitting a model to real data. Among those, Stochastic Block Models (SBM)~\cite{peixoto2019bayesian} are increasingly used to \emph{infer} a graph's block structure, whereby nodes of a given block are deemed to fulfill a similar role vis-à-vis nodes of other blocks -- this includes clusters, as well as more generic meso-level graph configurations, such as core-periphery~\cite{karrer2011stochastic}. 

Moreover, SBMs can be used to \emph{generate} synthetic graphs obeying specific behaviors~\cite{mangold2023}. 
For evaluation purposes, we thus rely on a recent SBM variant~\cite{mangold2023} enabling the parameterized joint generation of both cluster and core-periphery structures, to create a synthetic graph sequence $(S_0,...,S_{10})$.
Each phase consists of roughly 3-4,000 nodes, out of a potential of 11,000 (node labels), and obeys the following scenario: $S_0$ has two node groups, a core and a periphery; until $S_3$, group populations slightly change (i.e., some nodes enter or leave the graph, while others move from the periphery to the core); with $S_4$ and $S_5$ a higher proportion of periphery node population changes; a second periphery appears in $S_6$; with $S_7$, $75\%$ of each of the three node groups are renewed while preserving the meso-level structure; $S_8$ witnesses the emergence of a cluster; $S_9$ features again a $75\%$ renewal of each node group; finally with $S_{10}$, the population of each group slightly changes as was the case with $(S_1,...,S_5)$. On the whole, this scenario configures a variety of possible changes ---from conserving the graph structure while changing node labels, to changing the graph structure while keeping node labels--- in order to illustrate and validate the expected behavior of Chronoblox in such cases. 

For node group detection, we use both inferential and descriptive approaches to showcase the genericity of Chronoblox. We compute two meta-graph sequences: $\Sigma$ (using a minimal description length degree-corrected SBM unbiased towards cluster structures~\cite{peixoto2014efficient}) and $\Sigma'$ (using Louvain~\cite{blondel2008fast}). 

\begin{figure}[]
	\centering
	\includegraphics[width=\columnwidth]{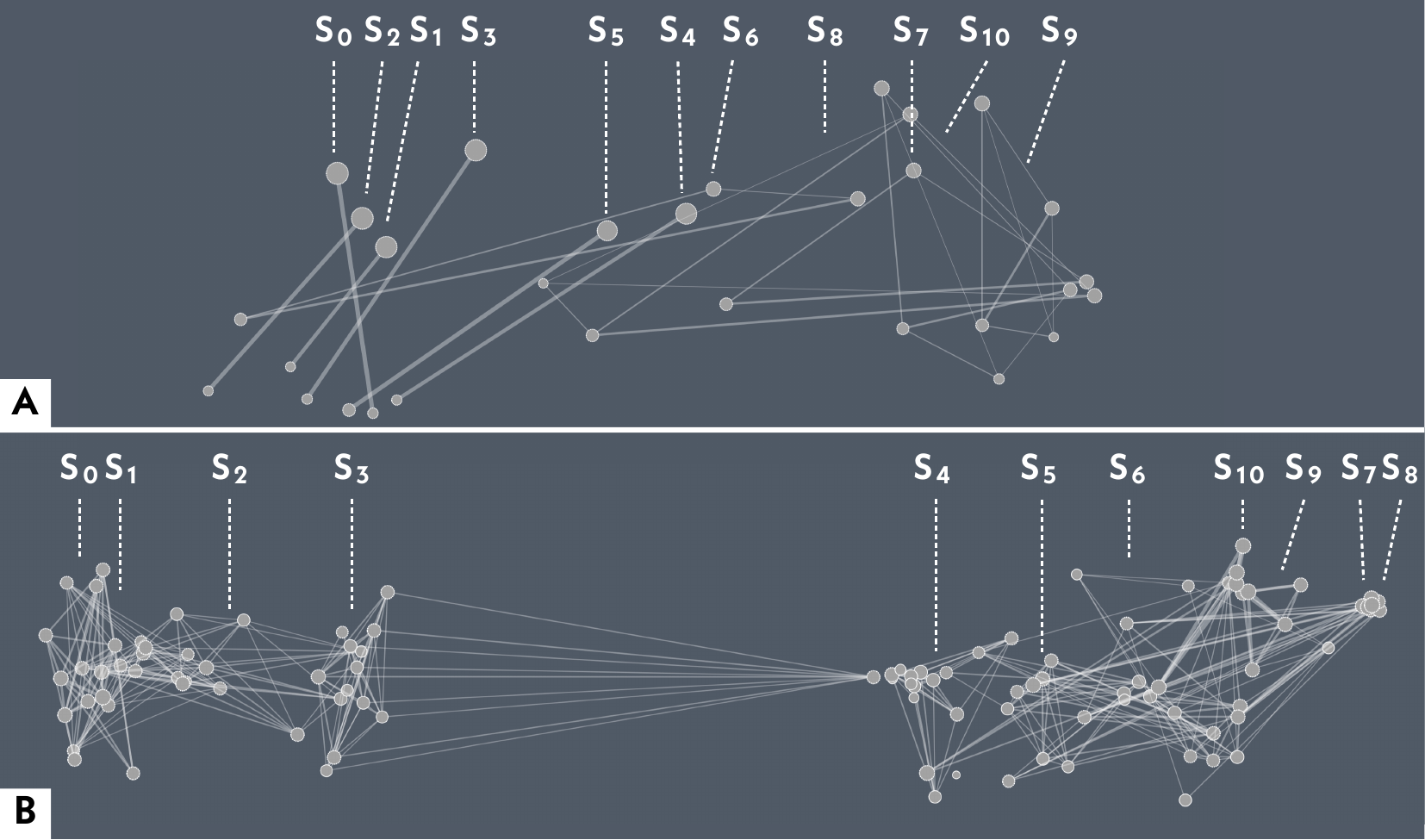}
	\caption{\textbf{Chronophotographies} of the synthetic graph sequences. \textbf{(A)} Nodes are grouped via the SBM method (\href{https://lobbeque.github.io/chronoblox_examples/toy_model_sbm.html}{interactive~version}), \textbf{(B)} Nodes are grouped via the Louvain algorithm (\href{https://lobbeque.github.io/chronoblox_examples/toy_model_louvain.html}{interactive~version}).}
	\label{fig:toy}
\end{figure}

\autoref{fig:toy}A shows the chronophotography of $\Sigma$. The initial peripheral group occupies the upper part of the visualization while the core is placed at the bottom. Chronoblox reflects the population changes from $S_0$ to $S_3$ with a slight shift from left to right. The rotation starting with $S_4$ conveys the turnover affecting the first peripheral group and also foreshadows the emergence of the second one with $S_6$. Significant population changes in $S_4$ to $S_6$ induce a macro-scale modification of the composition of the node groups of the corresponding $\Sigma$ meta-graphs, correctly resulting in a visible change in the areas where these newer meta-graphs are positioned. The subsequent major population turnover in $S_7$ triggers a pronounced jump to the right, with $S_8$ marking a further expansion to the left to integrate the fourth group. The second major turnover of $S_9$ once again leads to a big jump. Overall, Chronoblox manages to depict the transition from a core-periphery structure to a more assortative (clustered) system and to meaningfully translate graphically all the events described in its generative scenario. 

The chronophotography of $\Sigma'$, in \autoref{fig:toy}B, is less straightforward to interpret, for good reason too. Indeed, being less suited to the chosen core-periphery synthetic structure, Louvain breaks it down into a multitude of clusters that undergo changes which are weakly aligned with the graph overarching structure and the sequence scenario. We nonetheess still observe marked jumps that correspond to the population changes, first of $S_4$ and $S_5$, then of $S_7$ and $S_9$, with the meta-graphs occupying distinct areas. These differences highlight the importance of carefully selecting a node grouping approach which likely matches the underlying group structure of the graph, as they naturally affect the resulting chronophotography, eventually telling distinct evolutionary stories. As we shall now see with our empirical application, a Louvain-based node grouping approach would be more adapted to a sequence of graphs dominated by a cluster rather than a core-periphery structure. 

\section{Application}
\label{application}
We next use a real-world Twitter/X retweet network of users discussing the topic of ``\textit{Impact Investing}'' (II)~\cite{chiapello2020social}, based on a collection of 1.89M tweets posted between 2009 and 2021 using any term belonging to a set of II-related hashtags. 
After removing weakly active users (posting less than a tweet a month on average or less than a year apart), we infer likely home countries from geolocation shared in users' profiles and focus on the vast majority of users coming from English-speaking countries --- USA, UK, Australia and Canada. We end up with a dataset of 8267 users and over 100k timestamped links, that we divide into a sequence of yearly snapshots: $(X_{2009},...,X_{2021})$. We finally construct two sequence of meta-graphs using SBM $(\Xi_{2009},...,\Xi_{2021})$ and Louvain $({\Xi'_{2009}},...,{\Xi'_{2021}})$, shown respectively in \autoref{fig:interface} and \autoref{fig:teaser}.

This retweet network reflects an online counterpart of a social world that has historically been built up locally, whereby different versions of the II movement originated in the US and the UK and later circulated and spread country by country~\cite{chiapello2020social}. We can thus hypothesise geographical homophily to be a core element of the graph structure, reflecting the geographical consistency of these regional worlds. The chronophotography of $\Xi'$ in \autoref{fig:teaser}B appears to confirm this. First, each node group is colored according to the most frequent country of its nodes, which very often represents the majority as well: node clusters are geographically homogeneous, while enjoying a certain amount of connection with other clusters. Moreover, the curvilinear motion of the sequence of phase graphs reveals that these communities underwent a gradual population change --- in parallel, as the smooth rotation dynamics exhibits. Finally, the analysis of distances in the common visualization space yields further insights on the sequence. On the one hand, the lack of sudden jump indicates that this online, regionally fragmented world, has developed without major upheaval. On the other hand, the external arc, which is generally made of node groups dominated by the US (also the highest node population in the network), occupies and travels more space, hints at a further benefit whereby the resolution of the common embedding space increases when underlying node group populations are bigger and more diverse. 

The chronophotography of $\Xi$ additionally suggests that there is more at stake than just assortativity. For instance, the Canadian (orange) and British (light blue) node groups of \autoref{fig:interface}D exhibit the same core-periphery pattern that had been observed in $\Sigma$, evolving on parallel trajectories, belonging each to one lineage (one core, one periphery) that reflects the inter-temporal consistency and gradual change of both core and periphery populations, respectively. 

%
%

\section{Concluding remarks}
\label{conclusion}

The chronophotographies produced by Chronoblox show just how crucial the choice of a node grouping method is when it comes to visualizing dynamic graphs: each method reveals distinct and specific evolutionary processes. In this respect, one of Chronoblox's main strengths is that it is totally agnostic about the initial structural properties of the observed network, as well as those of the node groups shaped afterwards. This modularity allows Chronoblox to easily address a wide variety of research questions by swapping grouping methods or similarity measures at will. However, we would like to point out that Chronoblox cannot currently display more than a thousand node groups at a time. Future works will lead us to consider zooming and intra-temporal aggregation systems in order to visualize the evolution of larger or longer networks. 


\acknowledgments{
This work was supported by the ``Socsemics" Consolidator grant from the European Research Council (ERC) under the European Union's Horizon 2020 research and innovation program (grant agreement No. 772743)}

\bibliographystyle{abbrv-doi}

\bibliography{chronoblox}
\end{document}